\documentclass{elsart}

\usepackage{amssymb}
\usepackage{epsfig}

\begin{document}

\def\lapprox{{\raise0.5ex\hbox{$<$}\hskip-0.80em\lower0.5ex\hbox{$\sim$}
}}
\def\gapprox{{\raise0.5ex\hbox{$>$}\hskip-0.80em\lower0.5ex\hbox{$\sim$}
}}

\begin{frontmatter}

\title{Search for Narrow $NN \pi$ Resonances in Exclusive $pp \to
  pp\pi^+\pi^-$ Measurements}

\author[PIT]{W.~Brodowski}
\author[PIT]{J.~P\"atzold}
\author[PIT]{R.~Bilger}
\author[SLU]{H.~Cal\'en}
\author[PIT]{H.~Clement}
\author[SLU]{C.~Ekstr\"om}
\author[DRS]{K.~Fransson}
\author[HAM]{J.~Greiff}
\author[DRS]{S.~H\"aggstr\"om},
\author[DRS]{B.~H\"oistad}
\author[DRS]{J.~Johanson}
\author[DRS]{A.~Johansson}
\author[DRS]{T.~Johansson}
\author[IKP]{K.~Kilian}
\author[DRS]{S.~Kullander}
\author[SLU]{A.~Kup\'s\'c}
\author[DRS]{P.~Marciniewski}
\author[JIN]{B.~Morosov}
\author[IKP]{W.~Oelert}
\author[DRS]{R.J.M.Y.~Ruber}
\author[ITE]{M.~Schepkin}
\author[HAM]{W.~Scobel}
\author[SOL]{J.~Stepaniak}
\author[JIN]{A.~Sukhanov}
\author[IEP]{A.~Turowiecki}
\author[PIT]{G.J.~Wagner}
\author[IEP]{Z.~Wilhelmi}
\author[INS]{J.~Zabierowski}
\author[DRS]{J.~Zlomanczuk}

\address[PIT]{Physikalisches Institut der Universit\"at T\"ubingen, 
Morgenstelle 14, D-72076  T\"ubingen, Germany}
\address[SLU]{The Svedberg Laboratory, S-751 21 Uppsala, Sweden}
\address[DRS]{Department of Radiation Sciences, Uppsala University, S-751 21
  Uppsala, Sweden}
\address[HAM]{I. Institut f\"ur Experimentalphysik der Universit\"at Hamburg,
  Germany}
\address[IKP]{IKP - Forschungszentrum J\"ulich GmbH, D-52425 J\"ulich, Germany}
\address[JIN]{Joint Institute for Nuclear Research Dubna, 101000 Moscow,
  Russia}
\address[ITE]{Institute for Theoretical and Experimental Physics, 117218
  Moscow, Russia}
\address[SOL]{Soltan Institute for Nuclear Studies, PL-00681 Warsaw, Poland}
\address[IEP]{Institute of Experimental Physics, Warsaw University, PL-00681
  Warsaw, Poland}
\address[INS]{Soltan Institute for Nuclear Studies, PL-90137 L\'odz, Poland}
\begin{abstract}
Narrow structures in the range of a few MeV have been searched for in
$pp\pi^+$ and $pp\pi^-$ invariant mass spectra $M_{pp\pi^+}$ and
$M_{pp\pi^-}$ obtained from exclusive measurements of the $pp \to
pp\pi^+\pi^-$ reaction at $T_p = 725, 750$ and 775 MeV using the PROMICE/WASA
detector at CELSIUS.  The selected reaction is particularly well suited for the
search for $NN$ and/or $N\Delta$ decoupled dibaryon resonances.  
Except for a possible fluctuation at 2087 MeV/c$^2$ in $M_{pp\pi^-}$ no narrow
structures could be identified neither in $M_{pp\pi^+}$ nor in $M_{pp\pi^-}$
on the $3\sigma$ level of statistical significance,  giving an upper
limit (95\% C.L.)  for dibaryon production in this reaction of $\sigma < 20$ nb
for 2020 MeV/c$^2$ $< m_{dibaryon} < 2085$ MeV/c$^2$.
\end{abstract}

\begin{keyword}
Dibaryon, $NN\pi$ resonance
\PACS 14.20.Pb \sep 13.75.-n \sep 25.40.Ve
\end{keyword}
\end{frontmatter}

\section{Introduction}
With the realization of quarks  being the basic building
blocks of matter and of QCD being the appropriate theory of the strong
interaction, the idea emerged that the substructure of baryons should
lead to new types of states in the system of two baryons in addition to the
``trivial'' deuteron ground state and $^1$So virtual state. Indeed, QCD
inspired models in the early eighties \cite{1} predicted a large number of
dibaryon states of basic 6q structure. Despite a vast number of dedicated
experiments 
in search of such states  not a single one could yet be identified
unambiguously; for a review see, e.g. \cite{2}. However, the bulk of
these searches was devoted to dibaryons which by their quantum numbers could
couple to the $NN$ and/or $N\Delta$ channels. In these cases the dibaryon
states can undergo a fall-apart decay and hence cannot be  expected to have
narrow widths. Their widths should rather be even larger than those of
usual baryon resonances, which are in the order of 100 MeV and larger. Since
such very broad dibaryon resonances would be difficult to sense in
experiments, the failure of previous dibaryon searches could at least partly
be explained in this way.

On the other hand dibaryon resonances, which by their quantum numbers are
decoupled from the $NN$ and $N\Delta$ channels, may be expected to be rather
narrow, notably if close to the $NN\pi$
threshold. Actually, such states with $I(J^P) = 0(0^-)$ and $0(2^-)$ had
been predicted by Mulders et al. \cite{1} at a mass as low as $m = 2.11$
GeV/c$^2$  --- though  more recent theoretical work taking into account
proper antisymmetrization \cite{3} prefers the $0^-$ state at a somewhat higher
mass.

Recently a candidate for such a decoupled dibaryon state has been proposed for
the explanation of an otherwise peculiar, resonance-like structure in the
energy dependence of the forward angle cross section of the pionic double
charge exchange (DCX) on nuclei $A(\pi^+,\pi^-)B$ leading to discrete final
states \cite{4,5,6}. This structure, systematically observed on a wide range of
nuclei with peak cross sections at incident energies of $T_\pi \approx 40$ -
60 MeV, i.e. far below the $\Delta$ excitation, has been sucessfully explained
both in its energy and angular dependence by the assumption of the formation
of a narrow $NN\pi$ resonance, the so-called $d'$, with $I(J^P) =$ even
$(0^-)$, $m \approx 2.06$ GeV/c$^2$ and $\Gamma_{NN\pi} \approx 0.5$ MeV
\cite{4,5,6}. However, since this reaction takes place in the nuclear medium,
subtle medium effects cannot be excluded as origin of this structure. In fact,
there have been attempts to describe the observed 
resonance-like structures in special situations by  conventional models
\cite{7,8,9}, though a convincing description of the full observed systematics
is still lacking.

In order to minimize the effects of the nuclear medium, the DCX reaction has
also been carried out on $^3$He and $^4$He \cite{10}, i.e. on the lightest
nuclei, where 
this reaction is possible. However, contrary to the situation with heavier
nuclei, there is no longer a bound nuclear state in the exit channel, and the
process leads to the nuclear continuum only. Unfortunately, this situation
leads to a less pronounced signature of $d'$ production, in particular, if
collision damping of the $d'$ resonance with the neighbouring nucleons is
included \cite{10}. Hence, the DCX measurements on the He isotopes finally have
not been conclusive with regard to the existence of $d'$.

Also a search on the basic $NN$ system has been carried out at MAMI with a high
resolution and high statistics measurement \cite{11} of the reaction
$d(\gamma,\pi^0)X$. In the range 2020 MeV/c$^2 < m < 2100$ MeV/c$^2$ no narrow
structures had been found on the $3\sigma$ level with upper limits in the
range of a few microbarn  for the production of isoscalar or isovector
dibaryons . However, this limit is still an order of magnitude above the 
prediction for $d'$ production \cite{12} and hence not conclusive for this
particular dibaryon candidate either.

The only alternative reaction left in the basic $NN$ system and simultaneously
a potentially much more sensitive test of the $d'$ hypothesis is the reaction
$NN \to NN\pi\pi$, where $d'$ can be produced associatedly in a fixed target
experiment for projectile beam energies $T_p > 710$ MeV,  when assuming a mass
of $m_{d'} = 2.06$ GeV/c$^2$. The $pp \to pp\pi^+\pi^-$ reaction is
particularly well suited for this search, since both invariant mass spectra
$M_{pp\pi^+}$ and $M_{pp\pi^-}$ can be 
observed simultaneously. If a resonance shows up in both of these spectra,
then its isospin has to be $I=2$, if it shows up in $M_{pp\pi^-}$ only, then $I
\leq 1$. To this aim we have carried out exclusive measurements of the $pp \to
pp\pi^+\pi^-$ reaction at $T_p = 725, 750$ and 775 MeV using the PROMICE/WASA
setup \cite{13} with a hydrogen cluster jet target at the CELSIUS ring. The
ejectiles were detected in the angular range $4^\circ \leq \Theta_{lab} \leq
21^\circ$. Protons and pions were identified by the $\Delta E-E$ method,
$\pi^+$ particles were in addition positively identified by the delayed pulse
originating from $\mu^+$ 
decay following the $\pi^+$ decay at rest. This way the two-pion production
events could be clearly separated from the huge background of charged $\geq
3$-prong 
events due to single $\pi^0$ production with successive Dalitz decay or
$\gamma$ conversion into $e^+e^-$-pairs. As a test of the energy resolution we
constructed the missing mass spectrum from identified $pp\pi^+$ tracks and
obtained a clean single peak at the $\pi^-$-mass with a width of $\Gamma
\approx 8$ MeV FWHM  (see Fig. 1 in Ref. \cite{14}). The
four-momenta of the full $pp\pi^+\pi^-$ events were
reconstructed by kinematical fits with one over-constraint (1C). 
A total of 1163, 8016 and 9603 $pp\pi^+\pi^-$ events have been obtained at 725,
750 and 775 MeV, respectively.

>From Monte-Carlo (MC) simulations of the detector response we expect an energy
resolution of a few MeV in $M_{pp\pi^+}$ similar to that obtained for the
$\pi^-$ peak in the 3-particle missing mass $MM_{pp\pi^+}$. Note that in the
absence of a magnetic field  $\pi^-$
particles have not been detected, only reconstructed. Hence the resolution in
$M_{pp\pi^+}$ cannot be improved by using the single-particle missing mass
$MM_{\pi^-}$. 
For $M_{pp\pi^-}$ the resolution is significantly better, since this spectrum
being equivalent to the single-particle missing mass of the $\pi^+$ is
determined by  the four-momentum of the detected $\pi^+$ alone.
For the region of $d'$, i.e. near 2.06 GeV/c$^2$, we expect in
$M_{pp\pi^-}$  a resolution of FWHM = 3-4 MeV/c$^2$. Fig. 1 shows the
$M_{pp\pi^-}$ spectra obtained from the measurements at $T_p = 750$ and 775
MeV. These spectra, which are not yet corrected for efficiency and acceptance,
are compared to MC simulations including detector response. The dotted lines
show the result, assuming pure phase space for the reaction mechanism. The
shaded areas show the result of a model calculation, which quantitatively
reproduces all differential cross sections of the $pp \to pp\pi^+\pi^-$
reaction at $T_p = 750$ MeV \cite{15}. In this model the reaction is assumed
to proceed via the excitation of the $N^\ast(1440)$ resonance with its
subsequent decay into $N\sigma$ and $\Delta \pi$ channels. For a detailed
discussion of 
the reaction process see Ref. \cite{15}, where also the differential cross
sections for other invariant mass systems and angular distributions are
presented. Within statistics the data are smooth, compare very favorably with
the model calculations and, most importantly in the present context, show no
narrow structures on the $3\sigma$ level. To increase the sensitivity to $d'$
we next have imposed the condition $M_{pp} < 1896$ 
MeV/c$^2$ on events selected in $M_{pp\pi^-}$. This condition should enhance
the events stemming from $d'$ production relative to those from the
conventional process, since $d'$ --- if of dibaryonic nature --- should be
much smaller in volume than the interaction vertex of the conventional
process. Hence the protons originating from the $d'$ decay should undergo a
much stronger final state interaction, which in turn would lead to an
enhancement of events at low $M_{pp}$ masses \cite{16}. The thus constrained
$M_{pp\pi^-}$ spectra are shown at the bottom of Fig. 1. Again we do not
observe narrow structures of statistical significance with the possible
exception of a fluctuation at 2087 MeV/c$^2$, which however is  at the
high energy end of the accepted range and hence dangerous to interpret. For a
careful investigation of this excursion a measurement at some higher beam
energy would be necessary. There is also still some small enhancement
$(\lapprox~2\sigma)$ at 2.06 MeV/c$^2$ seen at the lower energy.

For $T_p = 725$ MeV the expected location of a possible $d'$ peak is already
very close to the high-energy end of the $M_{pp\pi^-}$ spectrum, where
kinematically the conventional 4-body phase space leads already to a peak-like
structure (see Fig. 2 in Ref. \cite{17}). In this situation the resolution in
$M_{pp\pi^-}$ and the accumulated statistics turned out to be too low for a
meaningful search for $d'$, while the situation is more favourable
at $T_p = 750$ and 775 MeV. 

On the basis of roughly 1000 events
obtained in a  test run \cite{14} at $T_p = 750$ MeV preceeding the high
statistics runs presented here  we had found a 4 MeV broad structure at 2.063
GeV/c$^2$ in $M_{pp\pi^-}$ with a statistical significance of $2\sigma$ to
$3\sigma$ depending on the treatment of the background. In this data analysis
only those $\pi^+$ events had been taken into account, for which a delayed
pulse from the secondary $\mu^+$ decay was  observed in the {\sl same}
detector element in which the $\pi^+$ particle was stopped. Because of the
layer structure of the 
PROMICE/WASA forward range hodoscope  this condition led to previously
unknown inefficiencies of $\pi^+$ detection for distinctive $\pi^+$ ranges,
and energies, respectively. This situation is reexamined in Fig. 2, where we
show the  $M_{pp\pi^-}$ spectra from the present work at 750 MeV 
with an order of magnitude more events than obtained in the test run. The
spectrum is now decomposed  
into events with a delayed pulse recorded in the same detector element where
the $\pi^+$ stopped (triangles) and events with a delayed pulse recorded in a
neighbouring element (open circles). Note that positrons originating from
$\mu^+$ decay  have a range in plastic detectors of up to 25 cm. Hence the
chance of finding a delayed pulse in a neighbouring element and simultaneously
none in the 
same element due to electronic thresholds is not negligible. We see that the
correction due to delayed pulses in the neighbouring elements is largest in
the mass range of 2065 to 2070 MeV/c$^2$ corresponding to the transition
region from the first to the second layer structure in the detector. The MC
simulations now correctly reproduce these 
features. Adding up the events with delayed pulses in the same and in the
neighbouring elements results in the solid dots, where some structure
at 2.063 GeV/c$^2$ still remains visible, however, strongly diminished as
compared to that in the  data denoted by triangles. The artifact nature of at
least part of 
the possible signal reported in Ref.~\cite{14} becomes obvious from this 
analysis. 

We note that in another measurement \cite{18} aiming to search for $d'$ in
this reaction also a bump near 2.06 GeV had been observed in $M_{pp\pi^-}$, if
a constraint on low $M_{pp}$ masses was imposed. These measurements conducted
at $T_p = 920$ MeV at ITEP had been carried out not on hydrogen but on C and
CH$_2$ targets. Hence the events of interest had to be obtained by subtracting
the C data from the CH$_2$ data. No follow-up studies have  been undertaken to
resolve the nature of that bump.

In Fig. 3 we present the $M_{pp\pi^+}$ spectra for $T_p = 750$ and 775
MeV. Again no structures of statistical significance are seen. Recall,
however, the lower energy resolution in $M_{pp\pi^+}$ which would wash out
narrow structures. In Fig. 4 we 
  finally show the differential cross sections for the  $M_{pp\pi^-}$ and
  $M_{pp\pi^+}$ distributions following corrections for efficiency and
  luminosity, and acceptance extrapolated to $4\pi$. From the observed 
  statistical fluctuations in these   data we  derive an upper limit (95\%
  C.L.) for the production cross section of narrow (FWHM $\lapprox$ 4 
  MeV) dibaryons of $\sigma_{dibaryon}~\lapprox~20 nb$. With respect to $d'$
  this upper limit is already more than one order of magnitude below the
  theoretical prediction \cite{16,19} of $\sigma_{d'} \approx 300$ - 1000 nb.  

In conclusion, we have carried out exclusive measurements of the $pp \to
pp\pi^+\pi^-$ reaction, which is particularly well suited for the search for
$NN$ decoupled dibaryon resonances. With the exception of the fluctuation
observed in $M_{pp\pi^-}$ at 2087 MeV/c$^2$, i.e. at the high energy end of
the covered range,  we find no narrow structures neither in $M_{pp\pi^+}$ nor
in $M_{pp\pi^-}$ within $3\sigma$ of statistical significance, which could be
indicative of dibaryon resonances in the mass 
range 2022 MeV/c$^2 < m < 2085$ MeV/c$^2$. We derive an upper limit 
of 20 nb for their production cross section in this reaction. In
particular for $d'$ 
this finding implies a number of consequences. Either it does not exist at
all, or its production cross section in this reaction is for some reason much
smaller than expected, or its mass outside the
nuclear medium is above the mass range investigated here,  or it exists only
in the presence of a nuclear medium. 

We acknowledge the continuous help of the TSL/ISV personnel and the support by
DFG (Graduiertenkolleg 683) and BMBF (06 TU 987).

\begin{figure*}
\begin{center}
\mbox{\epsfig{file=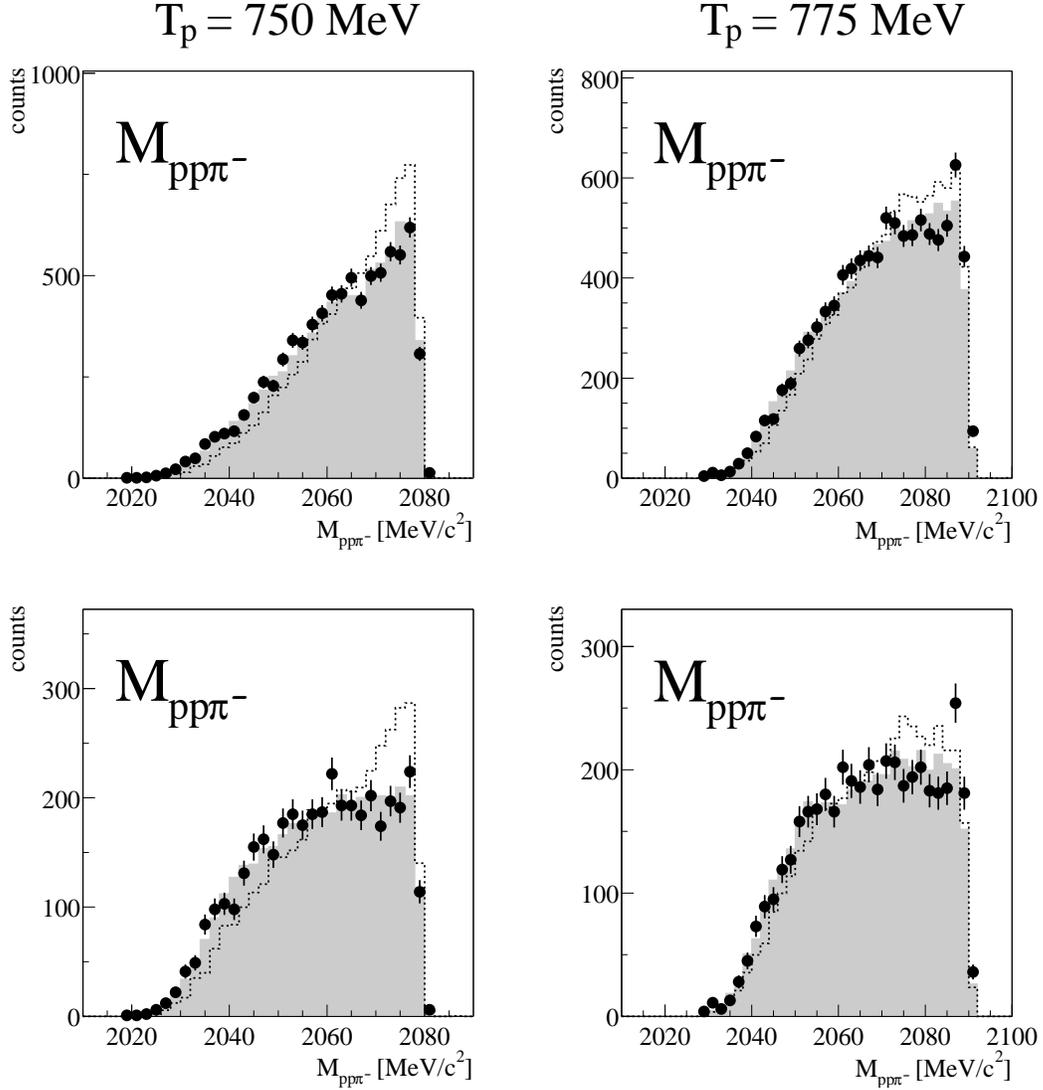, width=14cm}}
\end{center}
\caption{
Invariant mass spectra $M_{pp\pi^-}$ obtained from the exclusive
  measurements of the $pp \to pp\pi^+\pi^-$ reaction at $T_p = 750$ MeV (left)
  and $T_p = 775$ MeV (right). At the top the unconstrained spectra are shown,
  at the bottom the spectra contain only events meeting the condition $M_{pp}
  < 1896$ MeV/c$^2$ (see text). The dotted lines represent a MC simulation
  assuming pure   phase space for the reaction process, the shaded areas
  represent a model calculation (see Ref. \protect\cite{14}), which
  quantitatively describes all differential cross sections of the reaction. 
}
\end{figure*}

\vspace{0.5cm}

\begin{figure*}
\begin{center}
\mbox{\epsfig{file=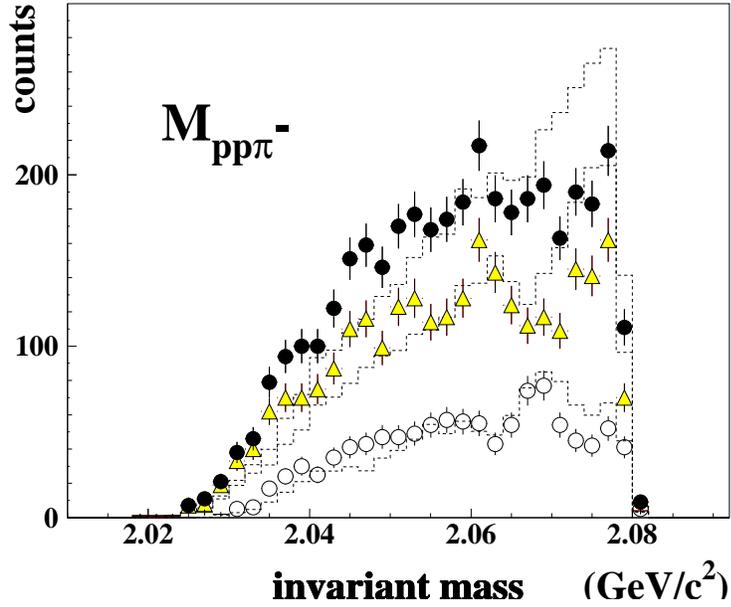, width= 10cm}}
\end{center}
\caption{
Invariant mass spectrum $M_{pp\pi^-}$ at $T_p = 750$ MeV. The
  triangles show events,  where the delayed pulse has been recorded in the
  same element where the $\pi^+$ particle stopped. The open circles show
  events, where the delayed pulse has been recorded by a neighbouring element
  only. The solid dots give the sum of both event classes and represent the
  full experimental result. The dotted histograms show MC simulations assuming
  pure phase space, but including $pp$ final state interaction (see ref.[14]).
}
\end{figure*}

\begin{figure*}
\begin{center}
\mbox{\epsfig{file=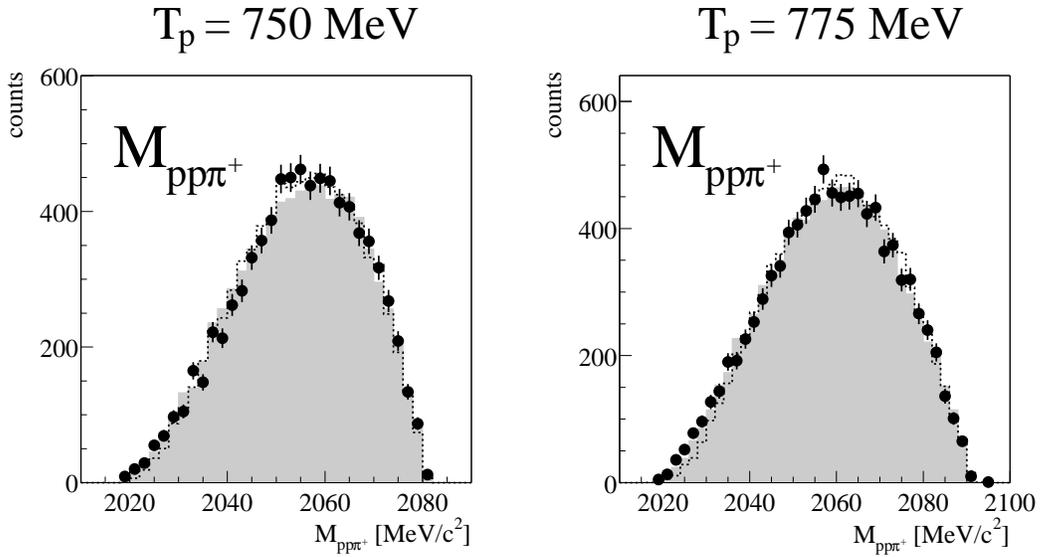, width=14cm}} 
\end{center}

\caption{
The same as Fig. 1, top, but for $M_{pp\pi^+}$. 
}
\end{figure*}

\vspace{0.5cm}

\begin{figure*}
\begin{center}
\mbox{\epsfig{file=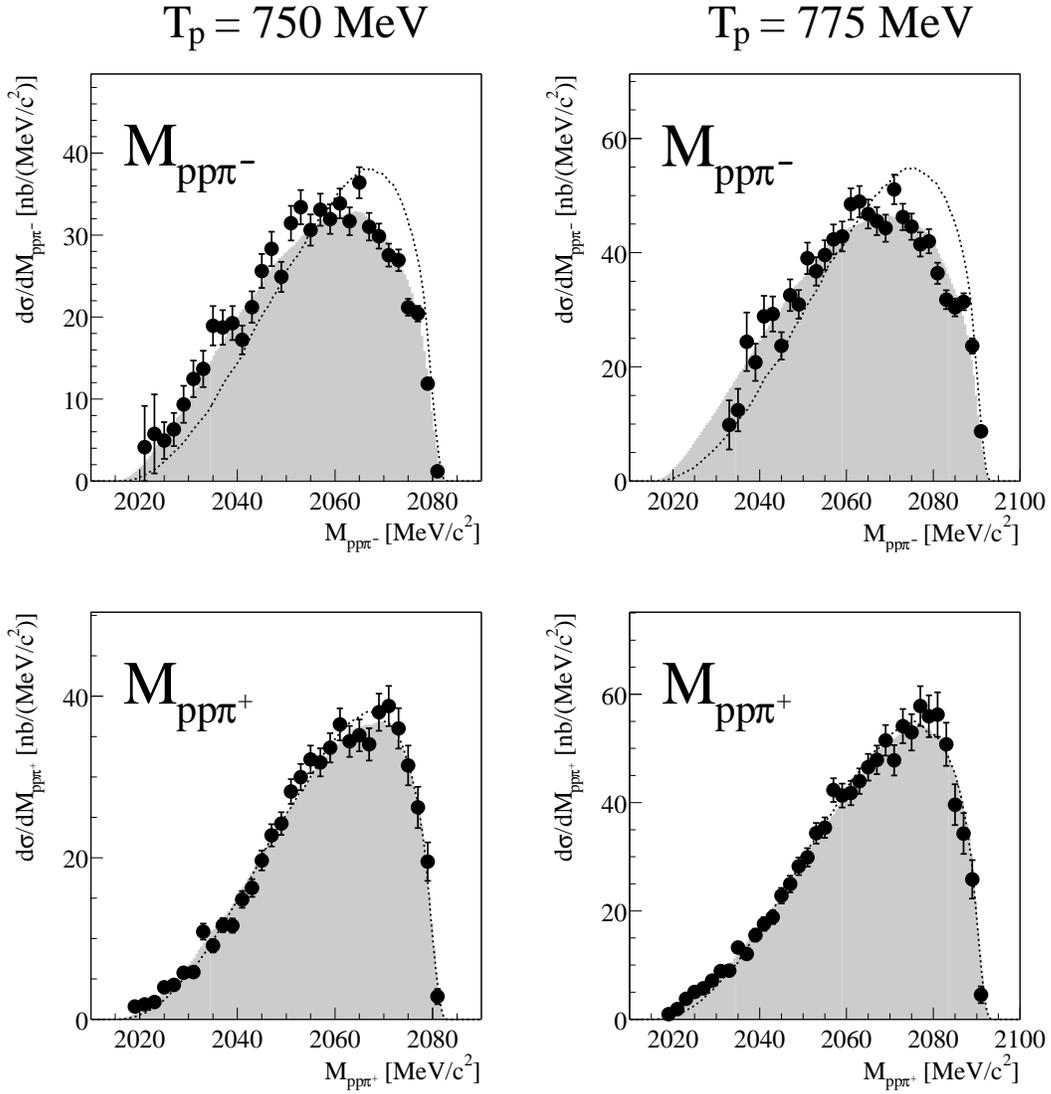, width= 14cm}}  
\end{center}

\caption{
Differential cross sections of the invariant mass distributions
  $M_{pp\pi^-}$ (top) and $M_{pp\pi^+}$ (bottom) at $T_p = 750$ MeV (left
  \protect\cite{14}) and
  $T_p = 775$ MeV (right). For the description of the curves see caption of
  Fig. 1.
}
\end{figure*}

\end{document}